\documentclass[prb,twocolumn,amsmath,floats,showpacs]{revtex4}
\usepackage{graphicx}% Include figure files
\usepackage{textcomp}

\begin{document}
\input epsf

\title{Reconstruction of the electron-phonon interaction function in Ta, 2H-NbSe$_{2}$
and MgB$_{2}$ from spectra of $S-c-N$ point-contacts}

\author{N.~L.~Bobrov \footnote {Email address: bobrov@ilt.kharkov.ua}, V.~V.~Fisun,
O.~E.~Kvitnitskaya, V.~N.~Chernobay, and \fbox{I.~K.~Yanson}}
\affiliation{B.I.~Verkin Institute for Low Temperature Physics and
Engineering, of the National Academy of Sciences
of Ukraine, prospekt Lenina, 47, Kharkov 61103, Ukraine}

\published {Fiz. Nizk. Temp. 2012, \textbf{38}, p.480(Low Temp. Phys. 2012, \textbf{38}, p.373)}
\date{\today}
%\(Submitted December 7, 2011)}
%\begin{comment*}[]
%\begin{center}
%{published in: Fiz. Nizk. Temp. 2012, \bold38, N5, pp.480-–492(Low Temp. Phys. 2012, \textbf 38, N5, pp.373--382; doi: 10.1063/1.4709437)}
%\end{comment}
%\end{center}

%Fiz. Nizk. Temp. 38, 480-492 (May 2012)

\begin{abstract}
Using Ta, 2H-NbSe$_2$ and MgB$_2$ as an example it is shown that it is possible to reconstruct
qualitatively a function of the electron-phonon interaction from point-contact spectra in a
superconducting state. The limits and the restrictions of this method are also shown. The
results obtained are compared with available literature data.

\pacs{71.38.-k, 73.40.Jn, 74.25.Kc, 74.45.+c, 74.50.+r.}
\end{abstract}

\maketitle

\section{Introduction}

A traditional point-contact (PC) spectroscopy of the
electron-phonon interaction (EPI) \cite{Kulik} assumes that
both electrodes forming a $N-c-N$-contact are in a normal
state at low temperatures. However, it is quite difficult to
transform many actual superconductors to a normal state.
Spectra of $S-c-N$ and $S-c-S$-point-contacts in some cases
contain an additional structure in the region of phonon
energies, and a form of such spectra is quite complicated
and on occasion is far from traditional. A transformation
into a normal state of electrodes forming a point-contact
often leads to disappearance of such a structure. It can
be explained by the fact that the spectrum in a normal state
and the additional nonlinearity arising at a transition to
a superconducting state are determined by scattering
processes in geometrically different regions of a
point-contact.

In a normal state the contribution to a spectrum is
{\em backscattering processes} at which an electron returns to
the same half space from which it flew out \cite{Kulik}. A main contribution
to a spectrum is formed by processes of scattering
of electrons in a volume with a diameter of the order of a
point-contact. Indeed, for a contact in the model of an aperture
with a diameter $d$ a probability of the return in the
case of the scattering is maximal in a center of the contact
and is 0.5. Since in an isotropic case a direction of the electron
flight after scattering is equally possible, on a large
distance $R\gg{}d$ from the contact plane the probability of
the return does not exceed a ratio of the contact area to the
surface area of a sphere with a center at the scattering point
and the radius $R$, and is $\sim$(d/4R)$^{2}$. Thus, on the distance
$R\sim2d$ this probability is $\sim32$ times lower than in the contact
center, and for $R\sim4d$ it is $\sim128$ times lower. A maximal
concentration of nonequilibrium phonons is achieved
in the region of maximal density of the current and
decreases fast with its spreading. So in the first approximation
at large distances from a narrowing the point-contact
can be considered as a point source of phonons, and their
density decreases $\sim1/R^2$. However, the largest part of phonons
is still generated in edges in the amount of the order
of the energy relaxation length of nonequilibrium electrons,
which depends on their energy. The inelastic mean
free path of electrons can be estimated by the formula
$l_{\varepsilon } =v_{F} \tau _{\varepsilon } $,
$$\tau _{\varepsilon }^{-1} =\frac{2\pi }{\hbar } \cdot \int _{0}^{eV}g(\omega)d\omega$$
here $g(\omega)$ is the Eliashberg thermodynamic function. Note
that if a metal has a few groups of electrons, then in the estimation
it is necessary to use the Fermi velocity for the group
which determines its transport characteristics. By the way,
this note is also valid for the expressions given in Sec. II for
an estimation of the potential barrier appearing at an interface
of two metals due to a mismatch of their Fermi parameters.
Obviously, that the scattering of nonequilibrium
electrons by phonons in edges does not affect the PC spectrum
in a normal state.
An inelastic superconducting addition in the spectrum is
determined by processes of scattering of electrons, undergoing
the Andreev reflection, by nonequilibrium phonons,
and {\em any scattering process} is effective. Since the probability
of reabsorption of nonequilibrium phonons by electrons
experiencing the Andreev reflection depends on the concentration
of phonons, from the above estimation it is clear that
the contribution to the superconducting inelastic spectrum
decays more slowly with increasing the distance from the
aperture than the contribution to the normal spectrum.
Besides, not taking into account phonons created in edges
would be too rough approximation, since the Andreev reflection
takes place in a volume of the order of the coherence
length, which is, in the cases under discussion, considerably
larger than the diameter of the point-contact, and is comparable
to the inelastic relaxation length at the Debye energies.
Estimations for the concentration of nonequilibrium phonons
depending on an applied voltage are reported in Ref.\onlinecite{Gloos} for a
Nb point-contact. It seems that a relative integral contribution
of the peripheral regions of a contact in the inelastic
superconducting addition in the spectrum for superconductors
with a sufficiently strong EPI can be comparable with
the contribution of the central regions. Therefore, if the contribution
to the spectrum of the point-contact central regions
is "turned off" somehow, the remained nonlinearity will
almost entirely be determined by the superconducting addition
to the spectrum of peripheral regions.
Taking into account that the impurities and defects disturbing
the crystal structure of the material under investigation
are usually concentrated near the surface, and considering that
in the process of forming point-contacts additional distortions
of the lattice are introduced, significantly lower quality of the
material near the contact center can be expected. This leads to
a blurring of the normal spectrum and a decrease in its intensity,
and to the complete suppression of it for a very strong
distortion of the lattice. At the same time edges of the contact
can remain quite clean, and the superconducting addition to
the spectrum will reflect properties of the bulk material. We
note in passing that the observed high level of the background
for some PC spectra in a normal state, which cannot be
described within the framework of the existing theory of reabsorption
of nonequilibrium phonons \cite{Kulik1}, possibly is also related
to the presence of a strongly distorted layer with a thickness
smaller than the diameter of the contact.
The aim of this work is reconstruction of the PC function
of the EPI from the superconducting addition to the
spectrum using clean point-contacts of Ta as an example.
Then, using algorithms obtained for Ta, the functions of the
EPI for 2H-NbSe$_2$ and MgB$_2$ are restored, spectra of which
in a normal state are absent or blurred very much.

\section{Main theoretical conceptions}
The authors of Ref.\onlinecite{Khlus} have extended a theory of the PC
spectroscopy to $S-c-S$ contacts. In this case it has been
assumed that $S-c-S$ contacts satisfy the following conditions:

\begin{equation} \label{GrindEQ__1_}
d\ll \zeta \qquad \frac{1}{\zeta } =\frac{1}{\xi (0)} +\frac{1}{l_{i} }
\end{equation}
\begin{equation} \label{GrindEQ__2_}
d\ll {v_{F} \mathord{\left/ {\vphantom {v_{F}  \omega _{D} }} \right. \kern-\nulldelimiterspace} \omega _{D} }
\end{equation}

These conditions mean that characteristic dimensions of the
contact are smaller than the momentum mean free path $l_i$
and the coherence length $\xi (0)$, the energy mean free path
$l_{\varepsilon } \gg d$ up to energies $\hbar \omega _D$,
($l_{\varepsilon }\sim {v_F\mathord{\left/ {\vphantom {v_F \omega _D }} \right. \kern-\nulldelimiterspace}\omega _D}$
for $\varepsilon \sim \hbar \omega _D $).
The $I-V$ characteristics (IVC) in the zero approximation in
${d\mathord{\left/ {\vphantom {d \zeta ,\quad {d\mathord{\left/ {\vphantom {d l_{\varepsilon}}}\right.\kern-\nulldelimiterspace} l_{\varepsilon }}}} \right. \kern-\nulldelimiterspace}\zeta,\quad {d\mathord{\left/ {\vphantom {d l_{\varepsilon } }} \right. \kern-\nulldelimiterspace} l_{\varepsilon } } } $
is given by the expression,
\begin{equation} \label{GrindEQ__3_}
I(V)=\frac{V}{R} +I_{exc}^{0}
\end{equation}
where
\textit{ $R^{-1} ={e^{2} p_{F}^{2} S\mathord{\left/ {\vphantom {e^{2} p_{F}^{2} S \pi ^{2} }} \right. \kern-\nulldelimiterspace} \pi ^{2} } $}
is the resistance in the $N$ state for a quadratic dispersion law; $S$ is the aperture area.
In the case of a real value of $\Delta $ for an excess current the result coincides
with that obtained in Ref.\onlinecite{Zaitsev}:
\begin{equation} \label{GrindEQ__4_}
I_{exc}^{0} =\frac{8\Delta }{3eR} th\frac{eV}{2T}
\end{equation}
Taking into account an effect of the EPI on a form of the
IVC at high $eV$ gives nonlinear in $V$ contributions to the current,
\[I(V)=\frac{V}{R} +\delta I_{ph}^{N} (V)+I_{exc}^{0} +\delta I_{ph}^{S} (V)\]
in this case, $\delta I_{ph}^{N} (V)$ coincides with a negative addition in the
current obtained for a normal contact \cite{Kulik},
\begin{equation} \label{GrindEQ__5_}
\delta I_{ph}^{N} (V)=-4eN(0)\Omega \int d\omega L(\omega ,eV,T)G^{N} (\omega )
\end{equation}
Here $\Omega\sim d^{3}$ is the volume of generation of phonons,
\begin{equation} \label{GrindEQ__6_}
L(\omega ,eV,T)=\frac{1}{2} \left(eV-\omega \right)\left[cth\frac{eV-\omega }{2T} +cth\frac{\omega }{2T} \right]
\end{equation}
The superconducting component of the inelastic current,
\begin{equation} \label{GrindEQ__7_}
\begin{array}{l} {\delta I_{ph}^{S} (V)=-4e\Omega N(0)\left[\int _{0}^{\infty }d\omega L_{1} (\omega ,eV,T,\Delta )G^{N} (\omega ) \right. +} \\ {+\left. \int _{0}^{\infty }d\omega L_{2} (\omega ,eV,T,\Delta )G^{S} (\omega ) \right]} \end{array}
\end{equation}
For $eV-\omega \gg \max (\Delta ,T)$
\begin{equation} \label{GrindEQ__8_}
L_{1} (\omega )=\frac{8\Delta }{3} \quad \quad \quad L_{2} (\omega )=\frac{2\Delta }{3}
\end{equation}

The function $G^{N} (\omega )$ in Eqs.\eqref{GrindEQ__5_} and \eqref{GrindEQ__7_}
is a PC function of the EPI which is similar to that in point-contacts in a normal state,
$G^{S} (\omega )$ is a superconducting PC function of the EPI
different from $G^{N} (\omega )$ by a form-factor. In this case, in contrast
to a normal form-factor causing the contribution to the current
due to electron-phonon collisions with changing the z-component
of electrons velocity, in the case of the superconducting
form-factor entering into $G^{S} (\omega )$ a contribution to the current
is made by electron-phonon collisions related to processes of
the type of the Andreev reflection in the contact region, i.e.
the transformation of quasielectron excitations into quasihole
ones. A relative value of the phonon contribution to the excess
current at $eV\sim\omega _D$ is of the order of
${d\cdot\omega _{D}\mathord{\left/{\vphantom{d\cdot\omega _{D} v_{F} }} \right.\kern-\nulldelimiterspace}v_{F} } $,
i.e. it is small if the condition \eqref{GrindEQ__2_} is fulfilled.
\noindent For the first derivative of the excess current it is obtained the relation
\begin{equation} \label{GrindEQ__9_}
\begin{array}{l} {\frac{dI_{exc} }{dV} =-\frac{64}{3R} \left(\frac{\Delta L}{\hbar \bar{v}} \right)\left[g^{N} (\omega )+\frac{1}{4} G^{S} (\omega )\right]} \\ {\quad \quad \quad \quad \quad \quad \quad \quad \quad \quad \quad \quad \quad \quad \quad \quad \omega ={eV\mathord{\left/ {\vphantom {eV \hbar }} \right. \kern-\nulldelimiterspace} \hbar } } \end{array}
\end{equation}

Since as shown in  Ref. \onlinecite{Khlus} at
$eV\ll\omega _D$, $G^{N}(\omega )\sim\left({\omega\mathord{\left/{\vphantom{\omega\omega _D}}\right.\kern-\nulldelimiterspace}\omega _D} \right)^4$, $G^{S}\sim\omega$,
then in the region of small displacements on the contact a main contribution is made by
processes without changing the z-component of the velocity, of the type of the
Andreev reflection.
In Ref. \onlinecite{Khlus1} $S-c-N$ point-contacts have been considered.
For the second derivative of the IVC it was obtained the following expression
\begin{equation} \label{GrindEQ__10_}
\begin{array}{l} {\frac{1}{R} \cdot \frac{dR}{dV} =\frac{16ed}{3\hbar } \cdot \sum _{a=1,2}\frac{1}{v_{F}^{(a)} }  \times } \\ {\times \int _{0}^{\infty }\frac{d\omega }{\Delta } \cdot S\left(\frac{\omega -eV}{\Delta } \right)G_{a} (\omega ) } \end{array}
\end{equation}
where $G_{a} (\omega )$ is the function of the EPI for normal and superconducting
metals forming a heterocontact, $S(x)$ is the blurring factor,
\begin{equation} \label{GrindEQ__11_}
S(x)=\theta (x-1)\frac{2\left(x-\sqrt{x-1} \right)}{\sqrt{x-1} }
\end{equation}
Thus, at $T\to 0$ a resolution is determined by $\Delta $. For the first derivative
of the excess current, the expression similar to the $S-c-S$ contact was obtained,
\begin{equation} \label{GrindEQ__12_}
\begin{array}{c} {\left. \frac{1}{R(V)} -\left(\frac{1}{R(V)} \right)\right|_{\Delta =0} =} \\ {=-\frac{32}{3R} \cdot \frac{d\Delta }{\hbar } \cdot \left[\frac{1}{v_{F}^{(1)} } \cdot G_{1} \left(\omega \right)+\frac{1}{v_{F}^{(2)} } \cdot G_{2} \left(\omega \right)\right]} \end{array}
\end{equation}
As it was mentioned above the expression for the excess
current of the pure $S-c-S$ contact, similar to \eqref{GrindEQ__4_}, was
obtained in Ref.\onlinecite{Khlus}. In that work as well as in Ref.\onlinecite{Artemenko}, a form
of the IVC for $S-c-S$ and $S-c-N$ contacts in pure and dirty
limits was found. In a dirty limit the excess current is lower
and at arbitrary temperature for the $S-c-S$ contact is
described by the expression,\cite{Artemenko}
\begin{equation} \label{GrindEQ__13_}
I_{exc} =\left. \frac{\Delta }{eR_{0} } \cdot \left(\frac{\pi ^{2} }{4} -1\right)\cdot th\frac{eV}{2kT} \right|\quad _{l_{i} \ll d}
\end{equation}
where $l_{i} $ is the momentum mean free path, $d$ is the contact diameter.
In the intermediate case, for an arbitrary relation
between $l_{i} $ and $d$ an evaluated relation, correct to first order
in $d/l_{i} $, was obtained in Ref.\onlinecite{Zaitsev1} for the $S-c-S$ contact,
\begin{equation} \label{GrindEQ__14_}
\frac{eI_{exc}^{(0)} R}{\Delta } =\frac{8}{3} \left(1-\frac{d}{5l_{i} } \right)
\end{equation}

Here \textbf{ }$R^{-1}=R_{0}^{-1}\left(1-d/l_i\right)$ is the contact resistance in a normal
state with taking into account the impurity scattering,
$R_{0}^{-1} =e^{2} SS_{F} \left\langle \alpha \right\rangle \left(2\pi \hbar \right)^{-3} $\textbf{\textit{- }}
is the resistance of an extremely pure contact. An absolute value of the excess current for
$S-c-N$ contacts appears to be twice smaller.
In Refs. \onlinecite{Khlus,Zaitsev,Khlus1}, upon calculating the excess current, it has
been assumed that the Fermi velocities and momenta coincide
at both edges of the point-contact. An effect of a barrier
within the plane of narrowing, appearing for different $v_F$ and
$p_F$, has been taken into account in Ref.\onlinecite{Zaitsev1}. For the excess current
of a pure $S-c-N$ heterocontact at arbitrary temperature
it was obtained the expression,

\begin{equation} \label{GrindEQ__15_}
I_{exc}^{(0)} =\frac{\Delta }{R_{N} } J{\rm th}(V/2T)
\end{equation}
\begin{equation} \label{GrindEQ__16_}
J=\frac{1}{2\left\langle \alpha D\right\rangle } \left\langle \alpha \frac{D^{2} }{R} \left[1-\frac{D^{2} }{\sqrt{R} } \cdot \frac{{\rm arcth}\sqrt{R} }{(1+R)} \right]\right\rangle
\end{equation}

Here $R=1-D$ is the reflection coefficient, $D=D(\alpha )$ is the
diffusion coefficient (calculation of $D(\alpha )$ and other coefficients
for heterocontacts is considered in detail in Ref. \onlinecite{Bobrov}).
For $D\to1$ $J=4/3$. With decreasing the diffusion coefficient
the form of the IVC and the dependence of the differential
conductivity $\delta =dI/dV$ on $V$ approach dependences typical
for tunneling contacts. The reflection from an interface
between metals leads to that $\delta (V)$ at low temperatures
becomes to be non-monotonic. At $T=0$,
\begin{equation} \label{GrindEQ__17_}
\left\langle \alpha D(\alpha )\right\rangle =\sigma _{N} /\sigma (\Delta )
\end{equation}
Thus the transparency can be estimated from experimental data.
For the $S-c-S$ contact for an arbitrary transparency of
the interface in the region of high voltages
\begin{equation} \label{GrindEQ__18_}
I_{exc}^{(0)} =\left(\Delta _{1} +\Delta _{2} \right)J/R_{N}
\end{equation}

where $J$ is determined by Eq. \eqref{GrindEQ__16_}.
As it was already mentioned above, the second derivative
of the IVC for the $S-c-N$ point-contact is given by the
convolution Eq. \eqref{GrindEQ__10_}. Taking into account the relation
between the derivative of the IVC and the PC function of
the EPI one can write down

\begin{equation} \label{GrindEQ__19_}
g_{pc}^{S} =\int _{0}^{\infty }\frac{d\omega }{\Delta } S\left(\frac{\omega -eV}{\Delta } \right)g_{pc}^{N} (\omega )
\end{equation}

\begin{figure}[t]
\includegraphics[width=8cm,angle=0]{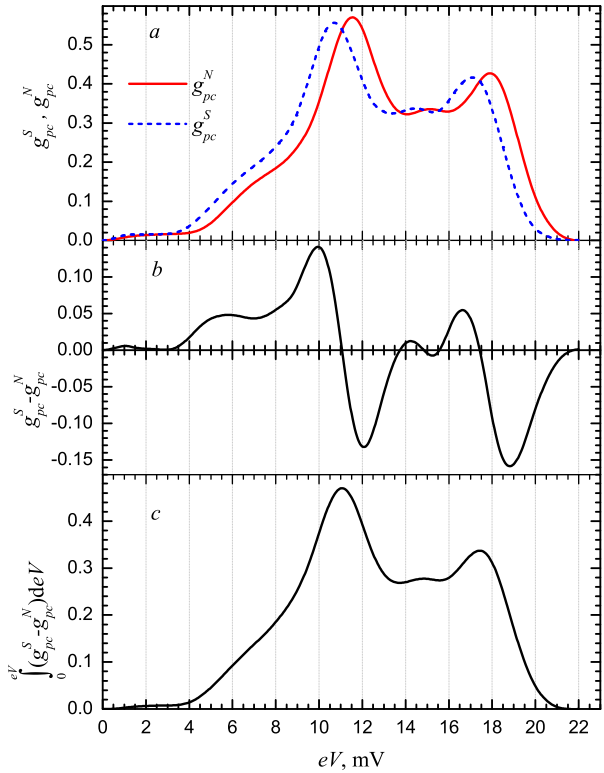}
\caption[]{(a) $g_{pc}^{N}$ - point-contact function of EPI in Ta, reconstructed from
the spectrum in Fig.\ref{FIG2}; $g_{pc}^{S}$ - a theoretically predicted transformation of the
point-contact function of EPI, upon a transition to a superconducting state
(Eq. \eqref{GrindEQ__9_}, see details in the text); (b) is a difference curve;
(c) is an integral from the difference curve, the PC function of EPI on the first
derivative of an excess current.}
\label{FIG1}
\end{figure}

As a model $g_{pc}^{N} (\omega )$ we will use a function of the EPI for
tantalum, restored from a spectrum of a heterocontact in a
normal state. Note that on writing the formula \eqref{GrindEQ__19_} we
neglected the fact that $G_{a} (\omega )$, corresponding to the superconducting
edge, is a superposition of \textit{g${}^{N}$($\omega $)} and \textit{ G${}^{S}$($\omega $)}
(see Eq.\eqref{GrindEQ__9_}). As it was mentioned above, \textit{G${}^{S}$($\omega $)}
is different from \textit{g${}^{N}$($\omega $)} by a form-factor, and most pronounced differences
are expected in the region of small displacements. In what follows,
since we obtained the spectrum of the EPI in a superconducting
state for the same contact, a comparison of
theory and experiment will be performed, and it will be considered
how much this approximation is acceptable.
The results calculated by Eq.\eqref{GrindEQ__19_} are shown in Fig.\ref{FIG1}.

In comparison with the initial curve $g_{pc}^{N}$, in the superconducting
curve $g_{pc}^{S}$ there is a shift of maxima of the EPI towards lower
energies by a value of the gap $\Delta $. (Note that in the $S-c-S$
contact for a transition into a superconducting state the position
of maxima of the EPI does not change.) Besides, its amplitude
is slightly smaller than that of the initial curve due to
an additional broadening on a blurring factor $S$ (Eq. \eqref{GrindEQ__11_}).
As it was already noted, in a superconducting state the spectrum
of the EPI should show up in the first derivative of the
excess current. Indeed, if from the superconducting curve $g_{pc}^{S}$
the initial normal curve $g_{pc}^{N}$ is subtracted, then we obtain the curve
$g_{pc}^{S} -g_{pc}^{N} $
(see Fig.\ref{FIG1}(b)). As it follows from Eq. \eqref{GrindEQ__12_},
the function of the EPI can be restored from the first derivative
of the excess current,

\begin{equation} \label{GrindEQ__20_}
g_{pc}^{S'} (eV)=\int _{0}^{eV}\left[g_{pc}^{S} (\omega )-g_{pc}^{N} (\omega )\right]d\omega
\end{equation}

This function is given by the curve shown in Fig.\ref{FIG1}(c).
Note that from Eq.\eqref{GrindEQ__9_} it follows that the point-contact function
of EPI in the first derivative of the excess current is a
superposition of $g^{N} (\omega )$ and $G^{S} (\omega )$. Thus, a position of phonon
maxima in $g_{pc}^{S'} (eV)$  is intermediate between
$g_{pc}^{S} $ and $g_{pc}^{N} $, that is seen in Fig.\ref{FIG1}.
Besides the inelastic component of the superconducting
addition in the spectrum, considered above, and taking only
place in contacts with the noticeable excess current, for the
point-contacts containing superconductors with a strong EPI
an elastic addition to the spectrum is also possible
\cite{Beloborod'ko,Omel'yanchuk}.
This addition is due to an energy dependence of the superconducting
order parameter $\Delta $($\varepsilon $) and takes place for both point-contacts
with direct conductivity (Eq.(10) in Ref.\onlinecite{Beloborod'ko}) and
tunneling point-contacts (Eq.(11) in Ref.\onlinecite{Beloborod'ko}). In Ref.\onlinecite{Omel'yanchuk}
it is noted that for ballistic contacts nonlinearity of the IVC of
the elastic origin can be comparable to inelastic point-contact
contributions. Obviously that for point-contacts with a
large tunneling parameter $Z$ it is exactly an elastic contribution
will be prevalent. For both tunneling and point-contacts
with direct conductivity the elastic addition to the spectrum
does not contain the function of the EPI g($\omega $) in explicit
form. However, it can be reconstructed by inverting the Eliashberg
equations (similar to the tunneling spectroscopy of
Rowell-McMillan \cite{Rowell}). For point-contacts with the direct conductivity
or small value of $Z$ it seems that the relation
between elastic and inelastic contributions is determined by
strength of the EPI. It is also necessary to take into account
that for different superconductors as the EPI increases the
energy mean-free path decreases leading to an increase in
the concentration of nonequilibrium phonons in the peripheral
regions adjacent to the contact (see, for instance, Ref.\onlinecite{Gloos}).
So for them one can expect an increase in a relative part
of nonlinearity, related to the inelastic superconducting scattering,
in comparison with superconductors with a weak coupling.
First experiments on measurements of point-contact
spectra of Sn and In, satisfying the conditions \eqref{GrindEQ__1_} and \eqref{GrindEQ__2_}
a superconducting state and their comparison with the spectra
in a normal state have been performed in Refs.\onlinecite{Yanson,Kamarchuk,Kamarchuk1,Kamarchuk2}.
The authors have found excellent agreement between the
Khlus theory \cite{Khlus1} and experimental results.
At the same time experiments in Pb (Ref.\onlinecite{Khotkevitch}) have
revealed a significant contribution of elastic processes. By
the way, in Fig.2 in Ref.\onlinecite{Khotkevitch} one can see a form of the elastic
contribution to the spectrum, calculated by the abovementioned
formula for the second derivative of the IVC of
the Pb $S-c-N$ point-contact, in comparison with experimental data.
In the Sec.III we will restore a function of the EPI
from the first derivative of the excess current assuming that
the elastic addition can be neglected. The point-contact spectra
of Ta, 2H-NbSe$_2$ and MgB$_2$ are used as objects of investigations.
We estimate a contact diameter by the Veksler interpolation formula,

\begin{equation} \label{GrindEQ__21_}
d\simeq \frac{\rho }{2R} +\sqrt{\left(\frac{\rho }{2R} \right)^{2} +\frac{16\rho l}{3\pi R} }
\end{equation}

For the sake of convenience we list parameters of the objects
under investigation:

\noindent Ta (Refs.\onlinecite{Wolf,Ryazanov,Poole}): -- T$_C$=4.47K; $\rho$l$\approx$5.9$\cdot$10$^{-12}${}$\Omega\cdot$cm$^{2}$; $\xi _0$=92nm; $\Delta_0$=0.71meV.

\noindent NbSe$_2$ (Refs.\onlinecite{Whitney,Takita}): -- T$_C$=7.2K; $\rho$l$_{ab}\approx$2.2$\cdot$10$^{-11}\Omega\cdot$cm$^2$; $\xi _{ab}$(0)=7.8nm; $\xi_{c}$(0)=2.6nm.
Values of the superconducting gaps within the two-band approximation (see, for instance,
Eqs.(2) and (3) in Ref.\onlinecite{Bobrov1}) estimated by us for the contacts given below with
the resistance of 1100 and 750 $\Omega$ respectively:
$\Delta_{1}$=1.1mV, $\Delta_{2}$=2.46mV, k=0.62, $\Delta_{aver}$=1.61mV; $\Delta_{1}$=1.07mV, $\Delta_{2}$=2.48mV, k=0.7, $\Delta _{aver}$=1.49mV; $Z$=0, $\Gamma_{1,2}$=0.

\noindent MgB$_2$ (Ref.\onlinecite{Prikhma}): -- T$_C$=39K; {$\rho$l$_{ab}$}$\approx$2.4$\cdot$10$^{-12}\Omega\cdot$cm$^2$; $\xi_{ab}$(0)=12nm, $\xi_c$(0)=3.6nm, $\Delta_{01}$=1.8$\div$2.4meV, $\Delta_{02}$=6.8$\div$7.4meV.

\section{Experimental Data}
\subsection{Point contacts Ta-Cu}
\begin{figure}
\includegraphics[width=8cm,angle=0]{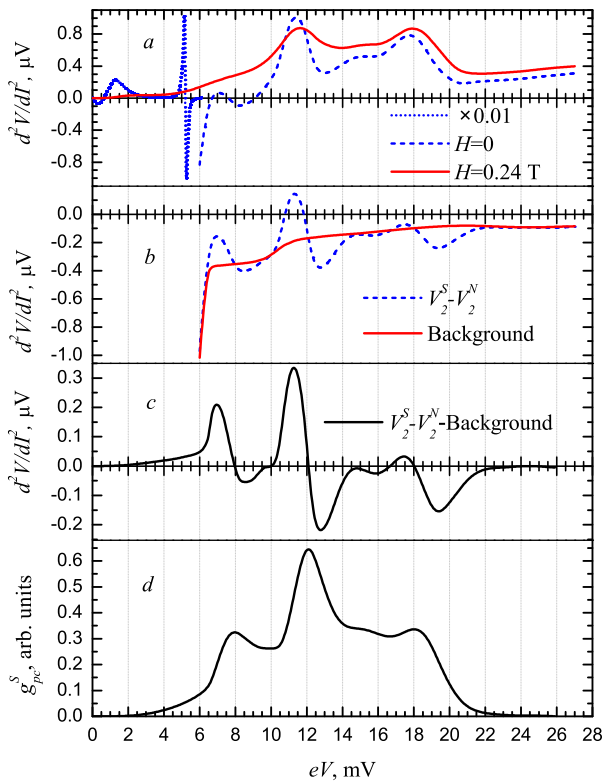}
\caption{(a)-Spectra of EPI of the Ta-Cu point contact in normal and
superconducting states: T=1.6~K, R$_{0}$=70~$\Omega$, an initial dashed part of the
superconducting curve containing the gap and nonequilibrium features is
normalized by 100; (b)-a difference between superconducting and normal
spectra, as well as an assumed form of a background curve; (c)-a difference
curve (after background subtraction); (d)-a point-contact function of
EPI reconstructed by integrating the previous curve.}
\label{FIG2}
\end{figure}

Fig.\ref{FIG2} shows the PC spectra of the contacts Ta-Cu in normal and superconducting states.

The resistivity of tantalum is
$\rho_{273K}$=12.6$\cdot$10$^{-6}\Omega\cdot$cm (Ref.\onlinecite{Startsev})
the ratio of resistances $\rho_{273K}$/$\rho_{res}\sim$18,
which gives an elastic mean free path 84~nm in the electrode at the helium temperature.
In this case the diameter of the point-contact with the resistance of
$\sim$70$\Omega$ is, according to Eq. \eqref{GrindEQ__21_}, 3.8~nm. Note that in the estimation
of the diameter the heterostructure was not taken into account.
However, considering that transport parameters of copper
({$\rho$l$\approx$}5.3$\cdot$10$^{-12}\Omega\cdot$cm$^2$ (Ref.\onlinecite{Gniwek}))
are close to those of tantalum, this correction can be neglected. A significantly
smaller background level in the superconducting spectrum is
to be noted. An identical decrease of background has been
noted for the first time in Ref.\onlinecite{Yanson}. In Ref.\onlinecite{Khotkevitch1} it has been
assumed that this effect can be related to decreasing the current
of quasiparticles upon combining some electrons responsible
for the excess current into Cooper pairs during the
Andreev reflection, since the EPI in the contact is connected
with a quasiparticle component of the total current. It is necessary
to note here that the shown spectrum satisfies the conditions
\eqref{GrindEQ__1_} and \eqref{GrindEQ__2_} and is similar to most intensive spectra
described in Ref.\onlinecite{Bobrov}. Nevertheless, upon a transition to a
superconducting state noticeable distinctions from predictions
of the theory are observed. Here we will not discuss an
appearance of nonequilibrium characteristics in spectra,
which have been considered in detail in Refs.\onlinecite{Bobrov2,Yanson1}, and
restrict our attention to the EPI. First of all we note that the
greatest distinctions are observed in the range of small energies.
A soft mode in the vicinity of 7$\div$8~mV starts to show up
as a peak. Instead of broadening, the peak close to
11$\div$12~mV becomes sharper abruptly, but the distinctions in
the vicinity of 18~mV are already not so pronounced. The
sharpening of the peaks can be caused by a high concentration
of nonequilibrium phonons with small group velocities
which leave the contact region slowly. Their velocities correspond
to the frequencies at which the density of states is
maximal. The probability of reabsorption of phonons by
electrons exhibiting the Andreev reflection increases with
the phonons concentration. It is just a selection of phonons
according to $d\omega /dq=0$  determines the effect of stabilization
of peaks positions and their sharpening upon a transition into
a superconducting state \cite{Yanson2}. In an $N$ state phonons participate
in inelastic backscattering processes in a broader area of the
phase space, and the position of maxima in the spectrum can
change depending on orientation of the contact axis with
respect to the crystallographic axes. In addition, the greatest
distinction of the curves in Figs. Fig.\ref{FIG1} and Fig.\ref{FIG2} exactly in the region
of small displacements can be due to the fact that an effect
of the superconducting form-factor entering to $G^{S} (\omega )$ was
not taken into account in calculating the convolution in Fig.\ref{FIG1}
(curve $g_{pc}^{S}$), see Eq. \eqref{GrindEQ__7_}. As it was mentioned in Introduction
it is just in the region of small displacements on the contact
an effect of the superconducting form-factor manifests itself
most pronouncedly, and the main contribution to the current
is made by processes of scattering without changing the $z$-component
of the velocity, of the type of the Andreev reflection.
The absence of the sharpening of phonon peaks in the
PC spectra of In and Sn upon a transition to a superconducting
state \cite{Yanson,Kamarchuk,Kamarchuk1,Kamarchuk2} is an indirect indication of a small contribution
of the region close to the contact in the inelastic superconducting
spectrum. For these superconductors the energy
relaxation length at energies of phonon peaks is higher than
that in Ta, and in spite of a small group velocity of such phonons
their concentration in edges in the vicinity of contractions
is insignificant to affect noticeably a form of the
spectrum. At the same time in the region of intensive generation
of phonons the selection on the basis of the phonon
group velocity is not so efficient due to a small size of the
current concentration region.
The presence in the $S$ spectrum of the gap and nonequilibrium
features as well as changing the form of the spectrum
and the reduced background makes it not possible to restore
the function of the EPI from the curve $V_{2}^{S} -V_{2}^{N} $
via simple integration. So, to compensate the influence of all neglected
factors before integrating it is necessary to subtract the background
as it is done for restoring the function of the EPI
from a spectrum of a normal state. Since at energies above
the Debye ones the function of the EPI turns to zero, \textbf{the
curve after subtracting the background should become to be
zero at these energies.} The second necessary condition is an
absence of a background on the curve obtained as a result of
integration. To fulfill this condition \textbf{the areas under the
curve after background subtraction should be the same
above and below an abscissa axis.} Note that above mentioned
requirements are satisfied for different forms of a
background curve. However, the variations do not lead to
significant changes of the form and positions of phonon
characteristics of a restored function of the EPI $g_{PC}^{S'} $
pc. So for a rough estimate of the EPI function, it is enough to plot a
background curve freehand. Because the gap and nonequilibrium
features do not allow to reconstruct correctly a trend
of the curve in the range from 0 to $\sim$6~mV, this part was
approximated by a parabola segment. In order to prevent
from additional misrepresentations the background curve
was smoothed out by segments of polynomial curves. We
note in passing that the restored function of the EPI $g_{PC}^{S'} $
is close enough in form to the EPI function of a superconducting
state, which could be obtained from the $S$--spectrum in
Fig.\ref{FIG2} (a)  by subtracting a traditional background.

\subsection{Point-contacts 2H-NbSe$_{2}$-Cu}

\begin{figure}
\includegraphics[width=8cm,angle=0]{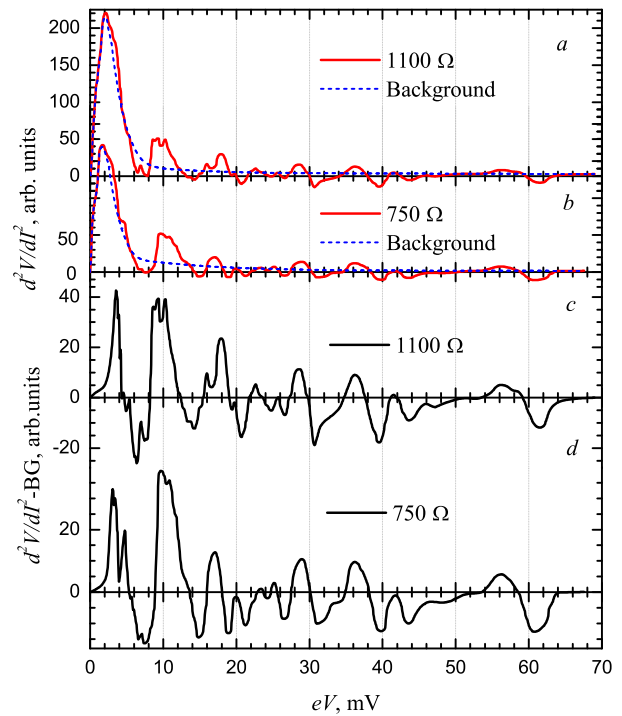}
\caption{(a) and (b) -- Point-contact spectra of 2H-NbSe$_2$-Cu and assumed
background curves, T=4.2~K, H=0. The contact with the resistance of
750$\Omega$ is obtained from the previous one by shortening. (c) and (d) --
Spectra after background subtraction}
\label{FIG3}
\end{figure}

The investigation of the EPI spectra of 2H-NbSe$_{2}$ in a
superconducting state is presented in Ref.\onlinecite{Bobrov3}. We emphasize
here that this compound is very difficult for experiments due
to a strong dependence of the point-contact spectra on the
quality of initial samples of single crystals. For example, one
of the first single crystals investigated by us had a resistance
ratio $\rho_{300}$/$\rho_{res}\sim$30. Point-contacts based on these single
crystals were characterized by a very low noise level up to
the resistance of 1 k$\Omega$. At the same time the point-contacts
based on single crystals with $\rho_{300}$/$\rho_{res}\sim$100 and having the
resistance higher than 10$\div$20$\Omega$, in most cases had a very
high level of noise that did not allow to use them for studying
the EPI. A possible reason for this is incommensurate
charge density waves (CDWs) that exist in this compound.
In the former case the single crystals are not perfect enough,
and there are effective pinning centers, preventing their
movement. In the later case there are not many of such centers,
and the motion of the CDWs in the vicinity of the contact
in the current state generates such noise.
Note a principal difference between the point-contact
EPI spectra in 2H-NbSe$_2$ Fig.\ref{FIG3} and Ta Fig.\ref{FIG2}.

\begin{figure}
\includegraphics[width=8cm,angle=0]{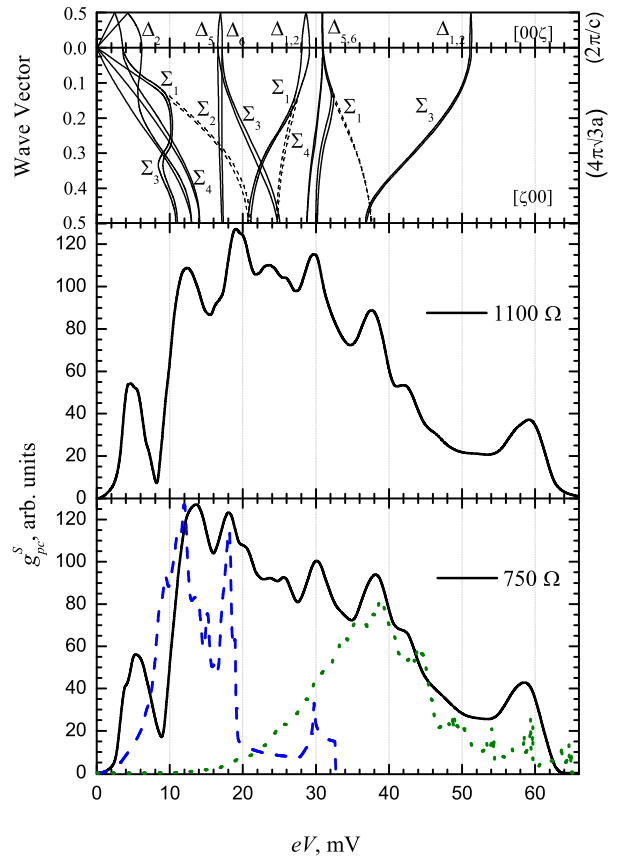}
\caption{A comparison of point-contact functions of EPI for 2H-NbSe$_2$,
reconstructed from the spectra (c) and (d) in Fig.\ref{FIG3} 3, with a function of the
phonon density of states from Refs.\onlinecite{Galetich} and \onlinecite{Gospodarev} (dash) as well as
with a twophonon convolution (dot) (the lower part of the figure) and the calculated dispersion
curves (model 1) from Ref.\onlinecite{Feldman} (for details see the text).}
\label{FIG4}
\end{figure}

In a normal state there is no structure in the region of phonon energies
in the spectra of 2H-NbSe$_2$ (see Fig.1 in Ref.\onlinecite{Bobrov3}).
At the same time in a superconducting state the spectra have many features.
It is noteworthy that by the standards of the traditional PC spectroscopy
contacts are very high-ohmic. The reasons for selecting such contacts are a
high resistivity of the compound, greater by about 2 orders of magnitude than in
conventional metals
($\rho_{\parallel}\sim$2$\cdot$10$^{-4}\Omega\cdot$cm; $\rho_{\bot}\sim$10$^{-3}\Omega\cdot$cm
(Ref.\onlinecite{Edwards})), and, apparently, a small energy mean free path of electrons
at energies close to the Debye due to the strong EPI.
The reason for the phonon features to appear only in a
superconducting state is apparently associated with a technology
of manufacturing the PCs. The sample of 2H-NbSe$_2$
was a single-crystalline plate with the dimensions of
$\sim$15$\times$20$\times$0.1~mm.
The electrodes with the dimensions
$\sim$3$\times$5~mm were cut out by a blade and glued with silver paste
to a wire holder. Thus, the contact pads of the electrodes
were machined, and a crystal structure of 2H-NbSe$_2$ on a site
of the cut was largely broken. Consequently, the point-contact
is very inhomogeneous; there are clean edges and a dirty
core with a strongly distorted lattice with a short mean free
path of electrons. In a normal state, an intensity of the phonon
spectrum is proportional to the ratio \textit{l$_i$/d} and very small.
The spectrum itself is formed by a material in a volume of
the order of a diffusion mean free path $\Lambda _{\varepsilon } $ and, because of
the crystal structure distortion, contains only a background
component. An estimate of diameters of the contacts by Eq.\eqref{GrindEQ__21_}
gives 2$\div$3.4~nm for the first and 2.3$\div$4.3~nm for the second
contacts respectively. An elastic mean free path for these
estimates was about 33 and 1.1~nm, corresponding in the former
case to the resistivity of unperturbed material in the vicinity
of T$_C$ and to $\rho$ in 2H-NbSe$_2$ at room temperature in the
latter case. In the other words, it was assumed that in the
contact region $\rho_{300}\rho_{res}\sim$1. Taking into account the lack of
spectra in a normal state, we can assume that the mean free
path near the narrowing is even smaller, and diameters of the
contacts are larger.
Upon a transition to a superconducting state at the IVC
of the point-contact there appears the excess current which
is formed by processes of the Andreev reflection at distances
of the order of the coherence length. This length is larger
than the diameter of the contact, and the greatest contribution
to the excess current is given by those of the edges
where the crystal lattice is the most perfect, because the
defects make worse superconducting parameters. Under this
assumption there is only dirty area of the superconducting
electrode, and the second normal electrode remains clean in
this case. This fact provides a unique opportunity to obtain
spectra that reflect a fine structure of the EPI of an undistorted
crystal lattice. As noted above, the superconducting
PC function of the EPI is proportional to a function of \textit{the
first derivative} of the excess current. Restoring the function
of the EPI from a superconducting spectrum in the absence
of any features on the second derivative in a normal state
assumes only subtraction of a monotonic background curve.
In this case, in the vicinity of the gap region ($\sim$3$\div$4~mV) the
background curves were plotted so that initial parts of the
difference curves (Fig.\ref{FIG3}, curves c, d) were parabolic.
The point-contact functions of the EPI restored from the
spectra c and d (Fig.\ref{FIG3}) via integration are shown in Fig.\ref{FIG4}.

Note that the available literature data on the Debye energy
of a phonon spectrum vary greatly among themselves. The
only known calculated function of the phonon density of
states, \cite{Galetich,Gospodarev}  according to the authors,
is based on elastic moduli and the maximum frequency of the quasiparticle mode
and is defined up to the energy of $\sim$32~meV. In this interval,
it agrees well with our data. One of the co-authors, \cite{Galetich,Gospodarev}
S.B.~Feodos'ev, suggested to associate the features observed at
higher energies with two-phonon processes, the calculated
function of which is shown in the same figure. The convolution
presented coincides perfectly in a form with the data
obtained by us, however so high intensity of two-phonon
processes is unclear to us. On the other hand, for much
higher frequencies the peaks observed on our curves are in
good agreement with flat parts of the calculated dispersion
curves, where $\partial\omega$/$\partial$q=0 in the Feldman theoretical model
for a dynamic matrix \cite{Feldman} (in the figure the first model is presented),
as well as with the calculations in Ref.\onlinecite{Motizuki}. The
exception is the most high-energy peak at $\sim$60~mV. Possible
causes of this peak are discussed in Ref.\onlinecite{Bobrov3}. A position of
the phonon features in our spectra correlates well with the
data of experimental works on the inelastic neutron scattering
and the Raman spectra. Unfortunately, the experimental
neutron data for this compound is generally limited to the
study of low-energy phonon branches. The small variations
observed in the position of the phonon features at $\sim$1$\div$2~mV
are apparently related both to an accuracy of the theoretical
model and an influence of the superconducting energy gap in
our data. Note that recently the point-contact functions of the
EPI have been obtained for the 2H-TaSe$_2$-compound belonging
to the same crystallographic group \cite{Kamarchuk3}. It is interesting that
a boundary of the phonon spectrum in this compound is
greater than 70~mV.

\subsection{Point-contacts MgB$_2$-Ag}

\begin{figure}
\includegraphics[width=8cm,angle=0]{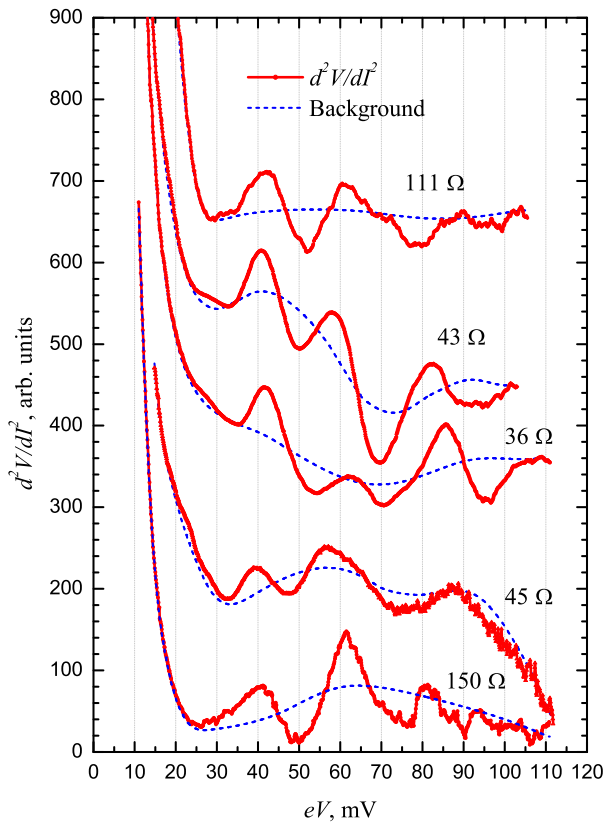}
\caption{Point-contact spectra of MgB$_{2}$ (T=4.2~K, H=0), taken from
Refs.\onlinecite{Yanson3,Yanson4,Yanson5}, as well as assumed background curves.
For clarity, the curves are shifted vertically.}
\label{FIG5}
\end{figure}

For the analysis, the superconducting PC spectra of
MgB$_2$ (Fig.\ref{FIG5}), reported in Refs.\onlinecite{Yanson3,Yanson4,Yanson5}, were used.

\begin{figure}
\includegraphics[width=8cm,angle=0]{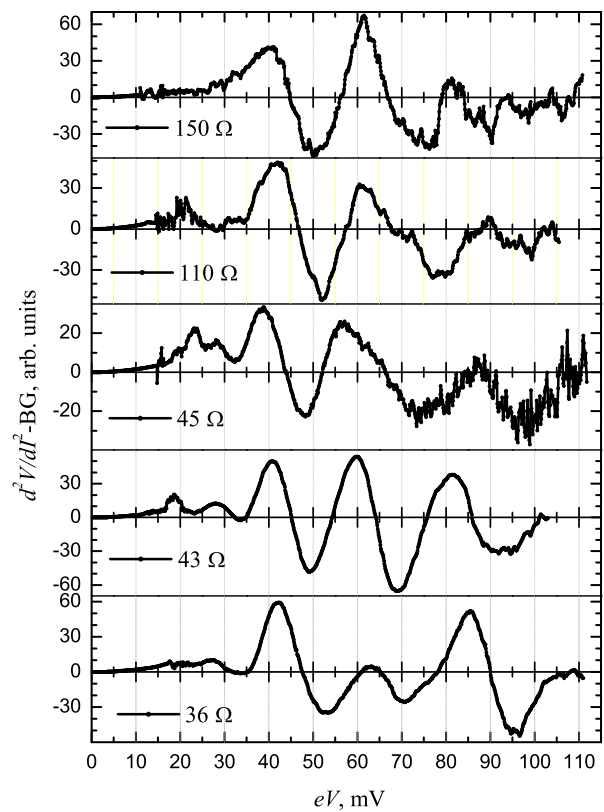}
\caption{Point-contact spectra of MgB$_2$ (Fig.\ref{FIG5}) after subtracting background.
Parts of the curve up to $\sim$18~mV are approximated by a parabola.}
\label{FIG6}
\end{figure}

The spectra were obtained by touching a massive silver electrode to
the film of MgB$_2$ and represent an intermediate case between
Ta and 2H-NbSe$_2$ for the appearance of features in a normal
state.  For the point-contacts similar to those shown in Fig.\ref{FIG5},
in the normal state the weakly pronounced highly blurred features are observed
in the region of the phonon energies with a high-level background \cite{Naidyuk}.
The residual resistivity of the film was 6$\cdot$10$^{-6}\Omega\cdot$cm,
that corresponds to the elastic mean free path of 4 nm.
The diameters of the given contacts are estimated to be 4.4,
3.9, 3.8, 2.2, and 1.9~nm (36, 43, 45, 110, 150 $\Omega$) respectively.
Thus, the estimates are very close to the estimated
diameter of 2H-NbSe$_2$, and the values of the elastic mean free
path in the contact region in the MgB$_2$ is found to be several
times larger. Such estimates are quite reasonable, considering
that we almost do not make additional lattice distortions.
When a background is subtracted form the curves in a superconducting
state, we were guided by the conditions specified
in SubSec.A and highlighted by bold. Fig.\ref{FIG5} shows the second
derivatives of the IVC and the background curves, and Fig.\ref{FIG6}
demonstrates the difference curves.

\begin{figure}
\includegraphics[width=8cm,angle=0]{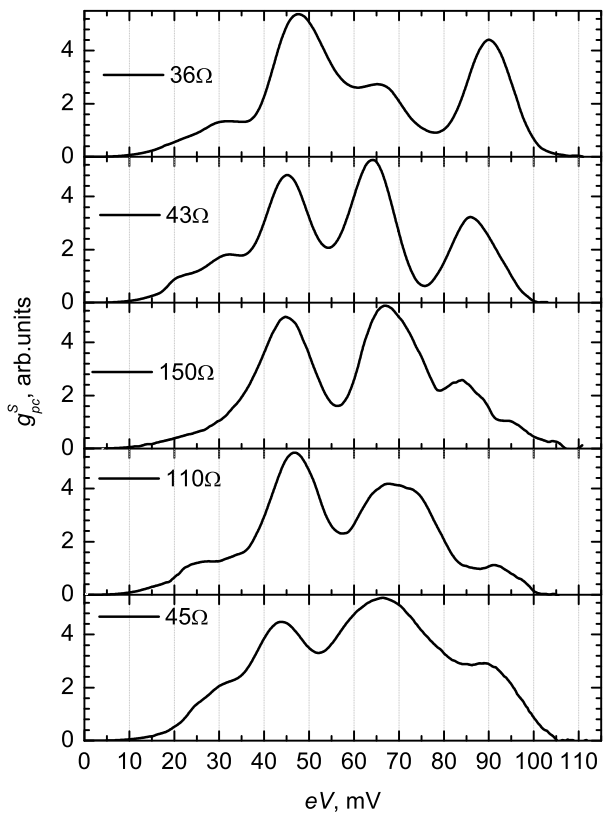}
\caption{Point-contact functions of EPI in MgB${}_{2}$, reconstructed from the difference
curves in Fig.\ref{FIG6}.}
\label{FIG7}
\end{figure}

As can be seen, the background curves are not monotonic as in the case of
2H-NbSe$_2$, and contain one or two blurred maxima. Note that
presented in Ref.\onlinecite{Naidyuk} point-contact spectra of the EPI in a normal
state as a whole are similar to the curves represented by
the background curves and contain the blurred maximum at
around $\sim$60~mV. The point-contact spectra of MgB$_2$ in a superconducting
state were interpreted in Ref.\onlinecite{Yanson5} from the point
of view of an energy dependence of the superconducting
order parameter $\Delta$($\varepsilon $).
The EPI function reconstructed from the difference
curves are shown in Fig.\ref{FIG7}.

\begin{figure}
\includegraphics[width=8cm,angle=0]{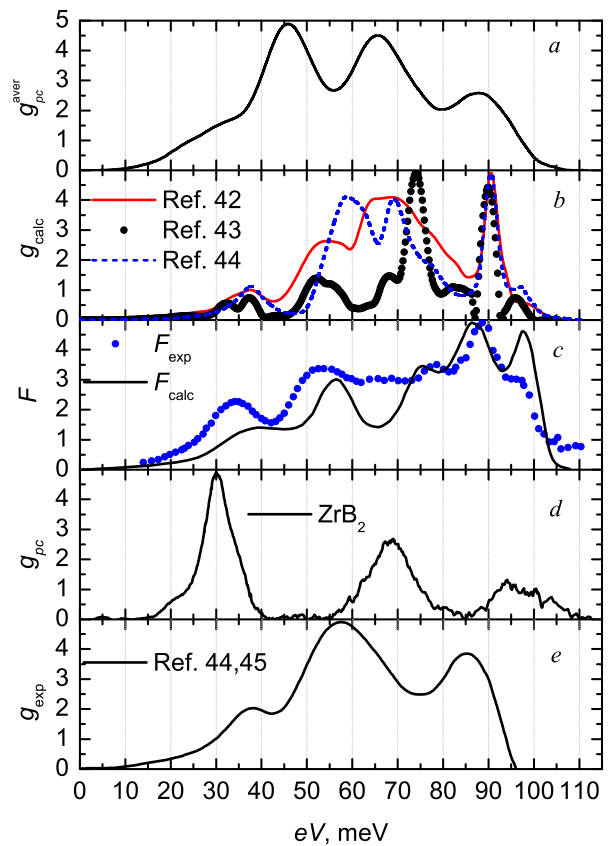}
\caption{A comparison of the point-contact function of EPI in MgB$_2$,
${\rm g}_{{\rm PC}}^{{\rm aver}}$ (a), averaged over five spectra (Fig.\ref{FIG7}):
with calculated functions of EPI, g$_{calc}$(eV) (Refs.\onlinecite{Pen,Golubov,Schneider}) (b);
functions of the phonon density of states (Ref.\onlinecite{Osborn}), calculated
F$_{calc}$(eV) and experimental F$_{exp}$(eV), obtained from neutron experiments (c);
the PC function of EPI gpc in ZrB$_2$ (Ref.\onlinecite{Naidyuk1}) (d);
the EPI function g$_{exp}$(eV) (Refs.\onlinecite{Schneider} and \onlinecite{Geerk}),
reconstructed from tunneling experiments (e).}
\label{FIG8}
\end{figure}

Fig.\ref{FIG8} demonstrates the PC function
of the EPI averaged over five contacts, ${\rm g}_{{\rm PC}}^{{\rm aver}}$,
related to it the calculated g$_{calc}$(eV)
(Refs.\onlinecite{Pen,Golubov,Schneider}) and obtained from the tunneling data
g$_{exp}$(eV) (Refs.\onlinecite{Schneider} and \onlinecite{Geerk}) functions of the
EPI as well as the functions of the phonon density of states
F$_{exp}$(eV) and F$_{calc}$(eV).\cite{Osborn}

Unfortunately, it is impossible to
compare in detail this function with the functions reconstructed
from the PC spectra in a normal state \cite{Naidyuk} due to a
rather diffuse structure of the latter. So we compare with the
point-contact function g$_{PC}$(eV) for ZrB$_2$ (Ref.\onlinecite{Naidyuk1}) belonging
to the same crystallographic group. As can be seen, the functions
are very similar in shape and position of the phonon
features which for the averaged spectrum are at the energies
by 4$\div$5~mV smaller than for ZrB$_2$, with the exception of the
first peak shifted to the higher energies by $\sim$16~mV. This is
understandable, since this mode is associated with vibrations
of the Zr ions, the atomic mass of which is $\sim$3.75 times greater
than that of the Mg ions. The calculated functions of the EPI g$_{calc}$(eV)
(Refs.\onlinecite{Pen,Golubov,Schneider} are notably different in form. However,
their comparison with the average PC function ${\rm g}_{{\rm PC}}^{{\rm aver}}$
demonstrates a close value in the energy of individual
branches of the spectrum. As for the tunneling function of
the EPI g$_{exp}$(eV),\cite{Schneider,Geerk} it can be assumed
that the maximum at 58~mV corresponds to the peaks at 46 and 66~mV of our averaged
functions. These peaks for the tunneling contact could
not be resolved due to insufficient quality of the superconducting
film. This assumption can also be supported by softening
the high-energy peaks from 90 to 85~mV. A
comparison of the function of the phonon density of states,
both calculated F$_{calc}$(eV) and obtained from the neutron data
F$_{exp}$(eV),\cite{Osborn} with our demonstrates a good correlation in a
position of phonon peaks. In general, it can be concluded
that the method used can accurately restore a form of the
EPI function from superconducting point-contact spectra.

\section{Discussion of the results}

As experiments demonstrate the method is effective for
metals both with a large coherence length and with rather
short one. For clarity, the characteristic parameters of the
point-contacts considered in this paper are given in the summary
table.

\begin{table*}[]
\begin{center}
\caption[]{Parameters of point-contacts Ta, NbSe$_2$, MgB$_2$.}
\begin{tabular}{|c|c|c|c|c|c|c|c|c|} \hline
 &\ $R,\; \Omega $ & $\rho l,\; \Omega \cdot {\rm cm}^{{\rm 2}} $ & ${\rho _{300} \mathord{\left/ {\vphantom {\rho _{300}  \rho _{res} }} \right. \kern-\nulldelimiterspace} \rho _{res} } $ & $\rho _{res}, \; \Omega \cdot {\rm cm}$ & $d,\; {\rm nm}$ & $l_{i} ,\; {\rm nm}$ & $\xi ,\; {\rm nm}$ & $\Delta ,\; {\rm mV}$ \\ \hline
Ta & 70 & 5.9$\cdot$10$^{-12}$ & 18 & 0.7$\cdot$10$^{-6}$ & 3.8 & 84 & 92 & 0.71 \\ \hline
NbSe$_2$ & 750$\div$1000 & 2.2$\cdot$10$^{-11}$ & 30 & 6.7$\cdot$10$^{-6}$ & 2$\div$2.3 & 33 & 7.7 & 1.1$\div$2.5 \\
& & & 1 & 2$\cdot$10$^{-4}$ & 3.4$\div$4.3 & 1.1 & & \\ \hline
MgB$_2$ & 36$\div$150 & 2.4$\cdot$10$^{-12}$ & 2.2 & 6$\cdot$10$^{-6}$ & 2.2$\div$4.4 & 4 & 12 & 2$\div$7 \\ \hline
\end{tabular}
\label{tab1}
\end{center}
\end{table*}

Based on the above material, we can conclude that an
interpretation of superconducting spectra is most straightforward
in two limiting cases:

\begin{enumerate}
\item for ballistic point-contacts in which the relative contribution
to the spectrum of a superconducting component, as
follows from the Khlus theory,\cite{Khlus1} is small and retains the
overall spectrum shape, sufficiently close to a normal state;
\item for point-contacts with a very dirty core and clean edges.
\end{enumerate}

Theoretically, a decrease of the elastic mean free path in
the contact region should lead only to a decrease in the intensity
of the spectrum and should not affect its shape. But in
reality, such a decrease of the mean free path is caused by either
impurities or defects and distortions of a crystal lattice.
In any case, this is a violation of short-range order, accompanied
by a blurring the spectrum. In the limiting case for
amorphization of a contact region the spectrum may be
absent completely, but the contact will not go into a heating
regime. If edges at the same time remain clean, after
subtracting a background the spectrum is easily obtained
from the first derivative of the excess current. Note that to
realize the situation of dirty narrowing and clean edges is
quite difficult. Most often, edges also appear to be with the
short-range order breaking.
A theory for the diffusion limit is presented in Ref.\onlinecite{Demin}.
The solution is obtained only for a long channel model
(bridge), provided that the bridge and the adjacent edge areas
contain large amounts of impurities. In the diffusion limit, as
in pure point-contacts, there are processes at which electrons
injected into a superconductor in the process of the Andreev
reflection interact with nonequilibrium phonons, giving rise
to a "back current." In addition, the contribution to the conductivity
is made by processes such as relaxation of charge
imbalance and the virtual processes that are sensitive to the
frequency dependence of the EPI function. In Ref.\onlinecite{Demin} it is
emphasized that these additional channels are observed only
for the geometry of a dirty bridge with a massive superconductor.
Obviously, in most cases, this geometry has little to
do with a real experimental situation. Departing from the
bridge geometry and considering the actual spatial dependence
of all quantities in the problem for a transition from a
narrowing to massive edges complicates significantly the
problem and still has not found a solution. Nevertheless, it is
obvious that a decrease of the elastic mean free path in the
vicinity of a narrowing reduces much faster a normal component
of the spectrum than superconducting. Therefore, upon
a transition from normal to superconducting state an intensity
of the spectra of the EPI increases in such point-contacts.
An example of such a transformation of spectra can be
observed in Refs.\onlinecite{Yanson6,Yanson7}.
An intermediate regime, close to diffusion one, is very
difficult to interpret. In this regime, both contributions are
similar in intensity, and the normal spectrum has a small
blur. A superposition of contributions of similar intensity
greatly distorts a shape of the resulting spectrum, and separation
of its superconducting part without carrying the contact
to a normal state is a highly non-trivial task \cite{Yanson8}. A complicating
factor is a possibility of appearing on the derivatives of
the CVC the nonspectral features of the thermal and nonequilibrium
origin, related to the destruction of superconductivity
in the contact region due to the spatial inhomogeneity
of a superconductor. To exclude from the consideration such
features there should be some statistics of the spectra, since
these features are not sufficiently reproducible. In addition,
in contrast to the spectral, their position on an energy axis,
are generally more dependent on temperature. A position of
the spectral features of the excess current on an energy axis
depends on the gap value and, hence, the temperature. However,
taking into account that the gap starts to decrease significantly
only at T$>$2/3T$_C$, in some temperature range this
position can be considered practically constant.
In addition to the above-noted difficulty of separating the
normal and superconducting contributions to the spectrum
and excluding the nonspectral features, the empirical method
of subtracting the superconducting background is rather arbitrary
yet. The problem of obtaining quantitative information
from the spectrum, especially in the case of irregular distribution
of impurities, is not solved. While in the simplest cases
for materials with a large coherence length, the purity of the
contact can be judged by a value of the excess current, in the
compounds with small $\xi$ the excess current is influenced by
other mechanisms in addition to the Andreev reflection.
Therefore, in complicated cases of the greatest practical interest,
such as high-temperature superconductors, this information
will only be of a provisional character.
Note that in the case of a short coherence length and a
relatively low critical current density in the current state, in
the vicinity of the point-contact there can be phase-slip surfaces
preventing the observation of phonon features \cite{Yanson9}.
Thus, though the PC superconducting spectroscopy is
not as universal as a traditional, in some cases it is more efficient.
First of all, these are compounds which are difficult to
take to a normal state, as well as compounds in which in a
normal state the spectra of the EPI are generally of poor
quality or absent at all, as, for example, in NbSe$_2$.

\section{Brief summary}

\begin{enumerate}
\item Using a Ta point-contact as an example it was shown that
it is possible in principle to reconstruct the point-conact
function of the EPI from the first derivative of the excess
current. It was found that the restored PC function of the
EPI differs in form from the EPI function of a normal
state, and the largest differences are observed at low energies.
An explanation of the observed differences was
proposed.

\item From the PC spectra in a superconducting state, for the
first time the function of the EPI was reconstructed in 2H-NbSe$_2$.
A position of phonon peaks in the EPI function
agrees well with the flat parts of calculated dispersion
curves.

\item The PC functions of the EPI in MgB$_2$, restored from
superconducting spectra for the first time, show good
reproducibility. A comparison with related functions of the
density of phonon states, F(x), tunneling and calculated
functions of the EPI shows a small shift of main phonon
peaks to lower energies in full agreement with theoretical
predictions.
\end{enumerate}

In conclusion, we wish to thank Yu.~G.~Naidyuk for
numerous tips, comments and additions made during discussion
of the paper. Our thanks are also to S.~B.~Feodos'ev for
a discussion of the phonon spectrum of 2H-NbSe$_2$.

\end{document}